\documentclass[pra, twocolumn, amsmath, amssymb, nofootinbib, floatfix]{revtex4}

\usepackage{graphicx, bm, float, hyperref}

\makeatletter
\def\graphicscale{\twocolumn@sw{0.3}{0.4}}
\def\graphicthreescale{\twocolumn@sw{0.3}{0.4}}

\begin{document}

\title{Measurement-induced dynamics of many-body systems at quantum criticality}

\author{Davide Rossini} 
\affiliation{Dipartimento di Fisica dell'Universit\`a di Pisa 
       and INFN Largo Pontecorvo 3, I-56127 Pisa, Italy}

\author{Ettore Vicari} 
\affiliation{Dipartimento di Fisica dell'Universit\`a di Pisa
       and INFN Largo Pontecorvo 3, I-56127 Pisa, Italy}

\date{\today}

\begin{abstract}
  We consider a dynamic protocol for quantum
  many-body systems, which enables to study the interplay between
  unitary Hamiltonian driving and random local projective measurements.
  While the unitary dynamics tends to increase entanglement, local
  measurements tend to disentangle, thus favoring decoherence.
  Close to a quantum transition where the system develops critical
   correlations with diverging length scales, the competition
  of the two drivings is analyzed within a dynamic scaling
  framework, allowing us to identify a regime (dynamic scaling limit) 
  where the two mechanisms develop a nontrivial interplay.  We
  perform a numerical analysis of this protocol in a measurement-driven
  Ising chain, which supports the scaling laws we put forward.
  The local measurement process generally tends to suppress quantum
  correlations, even in the dynamic scaling limit.
  The power law of the decay of the quantum correlations turns out
  to be enhanced at the quantum transition.
\end{abstract}

\maketitle

% ========================= BODY =========================

One of the greatest challenges of modern statistical mechanics is
understanding and controlling the quantum dynamics of many-body
systems.  The recent progress in atomic physics and quantum optical
technologies has provided a great opportunity for a thorough
investigation of the interplay between the coherent quantum dynamics
and the interaction with the environment, from both experimental and
theoretical viewpoints~\cite{MDPZ-12, HTK-12, RDBT-13, CC-13,
  Daley-14, SBD-16}.  The competition of such mechanisms may originate
a subtle interplay, likely representing the most intricate dynamic
regime of quantum systems where complex many-body phenomena may
appear. In this respect, it is worth focussing on situations close to
a quantum phase transition, where quantum critical fluctuations emerge
and correlations develop a diverging length
scale~\cite{Sachdev-book, SGCS-97}.

In general, while the unitary time evolution gives rise to a growth of
entanglement, measurements of observables disentangle degrees of
freedom and thus tend to decrease quantum correlations, similarly to
decoherence.  A quantum measurement is physically realized when the
interaction with a macroscopic classical object makes a quantum
mechanical system rapidly collapse into an eigenstate of a specific
operator, and the resulting time evolution appears to be a non-unitary
projection.  Such process is referred to as a projective
measurement~\cite{Zurek-03, vonNeumann-18}.  When the system is
projected into an eigenstate of a local operator, the corresponding
local degree of freedom is disentangled from the rest of the system.
Moreover, if measurements are performed frequently, the quantum state
gets localized in the Hilbert space near a trivial product state,
leading to the quantum Zeno effect~\cite{MS-77, FP-08}.

\begin{figure}[!b]
  \includegraphics[width=0.95\columnwidth]{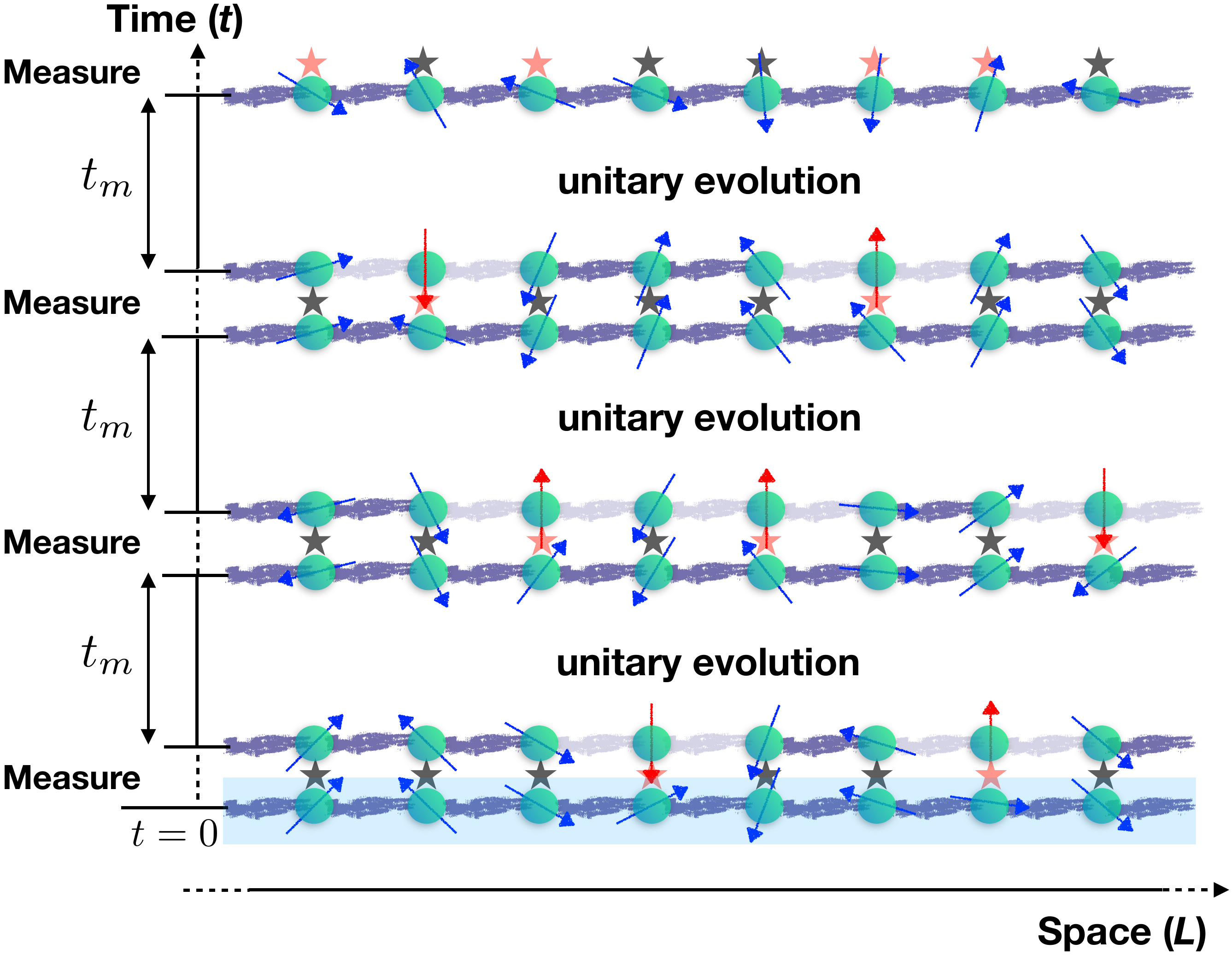}
  \caption{Sketch of the protocol: A quantum spin system, initially
    frozen in its ground state at quantum criticality ($t=0$), is
    perturbed with local projective measurements (stars) occurring
    after every time interval $t_m$, with a homogeneous probability
    $p$ per site. In between two measurement steps, the system evolves
    unitarily according to its Hamiltonian.  Red stars denote the
    occurrence of a measurement on a given site (for the sake of
    clarity, in the figure we considered $\sigma^{(3)}$-type measures:
    spins colored in red are projected along the $z$-axis).}
\label{sketch}
\end{figure}

Inspired by recent pioneering studies of the entanglement dynamics in
measurement-induced random unitary quantum circuits~\cite{LCF-18,
  CNPS-18, SRN-18}, we introduce a framework to address the interplay
of unitary and projective dynamics in experimentally viable many-body
systems at quantum transitions, such as quantum spin networks.  For
this purpose, we consider dynamic problems arising from protocols
combining the unitary Hamiltonian and local measurement drivings (for
a cartoon, see Fig.~\ref{sketch}).  In such conditions, it is not
clear how the presence of projective measurements modifies the quantum
critical behavior of a purely unitary system. One can easily imagine
that different regimes emerge, depending on the measurement protocols
and their parameters.  If every site were measured during each
projective step, then the system would be continually reset to a
tensor product state. A more intriguing scenario should hold when the
local measurements are spatially dilute.

Most of the work done so far in this context focused on the
investigation of entanglement transitions genuinely driven by local
measurements, either in random circuits~\cite{LCF-18, CNPS-18, SRN-18,
  LCF-19, SRS-19, GH-19, BCA-19, JYVL-19, GH-19-2, ZGWGHP-19}, or in the
Bose-Hubbard model~\cite{TZ-19, GD-20}, and recently on
measurement-induced state preparation~\cite{RCGG-19}.
In noninteracting models, continuous local measurements were shown
to largely suppress entanglement~\cite{CTL-19}. Here we focus
on a substantially different dynamic problem: understanding and
predicting the effects of local random measurements on the quantum
critical dynamics of many-body systems, i.e., when a quantum
transition is driven by the Hamiltonian parameters.

Specifically, we consider quantum lattice spin systems, assuming that
only one relevant Hamiltonian parameter can deviate from the critical
point, which we generically call $\mu$ and assume the critical point
to be at $\mu_c=0$, with corresponding renormalization-group (RG)
dimension $y_\mu>0$.  The system is initialized, at $t=0$, from the
ground state close to the critical point, thus $|\mu| \ll 1$. Random
local measurements are then performed at every time interval $t_m$,
such that each site has a (homogeneous) probability $p$ to be
measured.  In between two measurement steps, the system evolves
according to the unitary operator $e^{-i H(\mu) t}$, as sketched in
Fig.~\ref{sketch}, where $H(\mu)$ is its Hamiltonian and we fix
$\hslash = k_B = 1$.
If $p\to 1$, each spin gets measured every $t_m$, and the effects of
projections are expected to dominate over those of the unitary
evolution.  In contrast, for $p$ sufficiently small, the time
evolution may result unaffected by the measurements. In between these
two regimes, we unveil the existence of a competing unitary
vs.~projective dynamics, characterized by controllable dynamic scaling
behaviors associated with the universality class of the quantum
transition.

More complex protocols may be devised. For
example, the initial ground state might be replaced with a Gibbs state
at a finite temperature $T$. One may also consider a quench of
the control parameter at $t=0$, starting from the ground
state for a given value $\mu_0$ (so that $|\mu_0|\ll 1$), to a
different value $\mu$ which characterizes the unitary evolution
between the measurement steps.  In this case, the out-of-equilibrium
evolution arises from both the initial quench and the measurement
protocol.  For the sake of clarity in our presentation, we will focus
on the simpler version discussed before, even though an extension to
such more complex scenarios is not difficult (see Appendix~\ref{phedynscathe}).

As for the model, we consider the paradigmatic $d$-dimensional quantum
Ising Hamiltonian,
\begin{equation}
  H_{\rm Is} = - J \, \sum_{\langle {\bm x}, {\bm y}\rangle} \sigma^{(3)}_{\bm x} \sigma^{(3)}_{\bm y} 
  - g\, \sum_{\bm x} \sigma^{(1)}_{\bm x}  
  - h \,\sum_{\bm x} \sigma^{(3)}_{\bm x} \, ,
  \label{hedef}
\end{equation}
where ${\bm \sigma}\equiv (\sigma^{(1)},\sigma^{(2)},\sigma^{(3)})$
are the spin-1/2 Pauli matrices, the first sum is over the bonds
connecting nearest-neighbor sites $\langle {\bm x}, {\bm y}\rangle$, while the
other sums are over the sites.  We fix $J=1$ as the energy scale.  At
$g=g_c$ and $h=0$, the model undergoes a continuous quantum transition
belonging to the two-dimensional Ising universality class, separating
a disordered phase ($g>g_c$) from an ordered ($g<g_c$)
one~\cite{SGCS-97, Sachdev-book}.  Such transition is characterized by
a diverging length scale $\xi$ of the critical correlations, and the
suppression of the energy gap $\Delta$ as $\Delta \approx \xi^{-z}$,
where $z=1$ is the dynamic exponent.  The power-law divergence of
$\xi$ is related to the RG dimensions of the relevant parameters
$\delta \equiv g-g_c$ and $h$: it behaves as
$\xi\sim |\delta|^{-1/y_\delta}$ at $h=0$, and $\xi\sim |h|^{-1/y_h}$ for
$\delta=0$~\cite{Ising1D}.

In our dynamic protocol, we take a spin system of linear size $L$ with
periodic boundary conditions and perform, on each site, local random
measurements of the spin components ${\bm \sigma}_{\bm x}$,
along the transverse [$\sigma_{\bm x}^{(1)}$] or the
longitudinal [$\sigma_{\bm x}^{(3)}$] directions, every time interval
$t_m$ and with probability $p$.  We then project onto the measured
value of the spin component, and normalize the many-body wave
function.  The main features of the resulting evolution are inferred
by fixed-time averages of observables, such as magnetization $m(t)$
and susceptibility $\chi(t)$:
\begin{equation}
  m(t) = \frac{1}{L^d} \sum_{\bm x} \langle \sigma_{\bm x}^{(3)} \rangle_t \,, \quad
  \chi(t) = \frac{1}{L^d} \sum_{{\bm x},{\bm y}} \langle \sigma_{\bm x}^{(3)} \sigma_{\bm y}^{(3)} \rangle_t \,,
\end{equation}
averaging over trajectories ($\langle \cdot \rangle_t$ is
the expectation value of observables at time $t$).
Since measurements generally suppress
quantum correlations and $\chi(t\to\infty) \to 1$, corresponding to an
uncorrelated state, to monitor this process we study the ratio
$R_\chi(t)\equiv [\chi(t)-1]/[\chi(t=0)-1]$, which goes
from one ($t=0$) to zero ($t\to\infty$).
The time scale $\tau_m$ of the suppression of quantum correlations
may be estimated from the halving time of $R_\chi(t)$.

\begin{figure}[!t]
  \includegraphics[width=0.95\columnwidth]{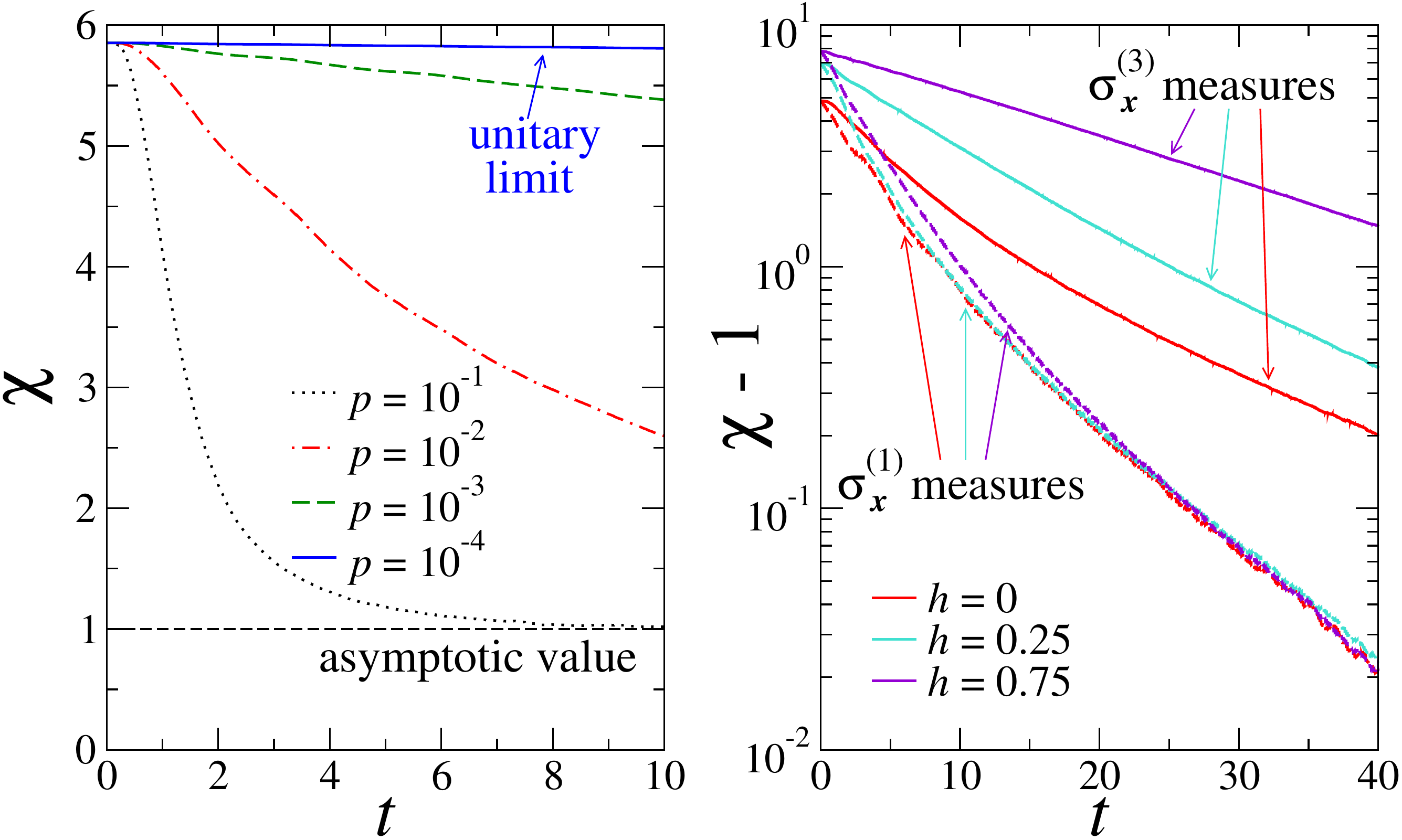}
  \caption{Evolution of the susceptibility $\chi$ for the Ising chain
    with $L=10$ spins and $g=1$.  Local measurements are performed
    along either the longitudinal direction [$\sigma^{(3)}_x$,
      continuous curves] or the transverse direction
    [$\sigma^{(1)}_x$, dashed curves], with a time interval $t_m =
    0.1$.  In the left panel we fix $h=0$ and vary $p$; in the right
    panel we fix $p=10^{-2}$ and vary $h$, as indicated in the
    legends.  Here and in the next figures data are averaged 
    over many [$O(10^4)$] trajectories and errors are of the orders of
    the size of the lines.}
\label{basic}
\end{figure}

As visible from the data in Fig.~\ref{basic}, obtained by numerically
simulating the dynamic protocol for a one-dimensional (1D) quantum
Ising model~\cite{Numerics}, the random and local spin projections
tend to destroy correlations in the system, which converges
asymptotically in time to a fully disordered configuration ($\chi=1$,
$m=0$).  The time scale $\tau_m$ of such dynamical process depends on
$p$ (left panel), as well as on the initial state and the measurement
axis (right panel).  In particular, longitudinal-field measurements
are less destructive than those along the transverse field, being
orthogonal to the coupling and thus to the ordering direction of
spins.
One can gain more insight on this mechanism, which resembles a
relaxation process due to decoherence, by first looking at the exactly
solvable single-spin model (see Appendix~\ref{onespin}, which contains
the analytic solution of that model): Irrespective of the magnetic
field strength and direction, for finite $t_m$, the spin magnetization
drops to zero exponentially in time, ending up into a completely
unpolarized state.  Looking again at the decay highlighted by the
curves in Fig.~\ref{basic}, deviations from pure exponentials have
thus to be ascribed to the full many-body nature of the
system~\eqref{hedef}.

\begin{figure}[!t]
  \includegraphics[width=0.95\columnwidth]{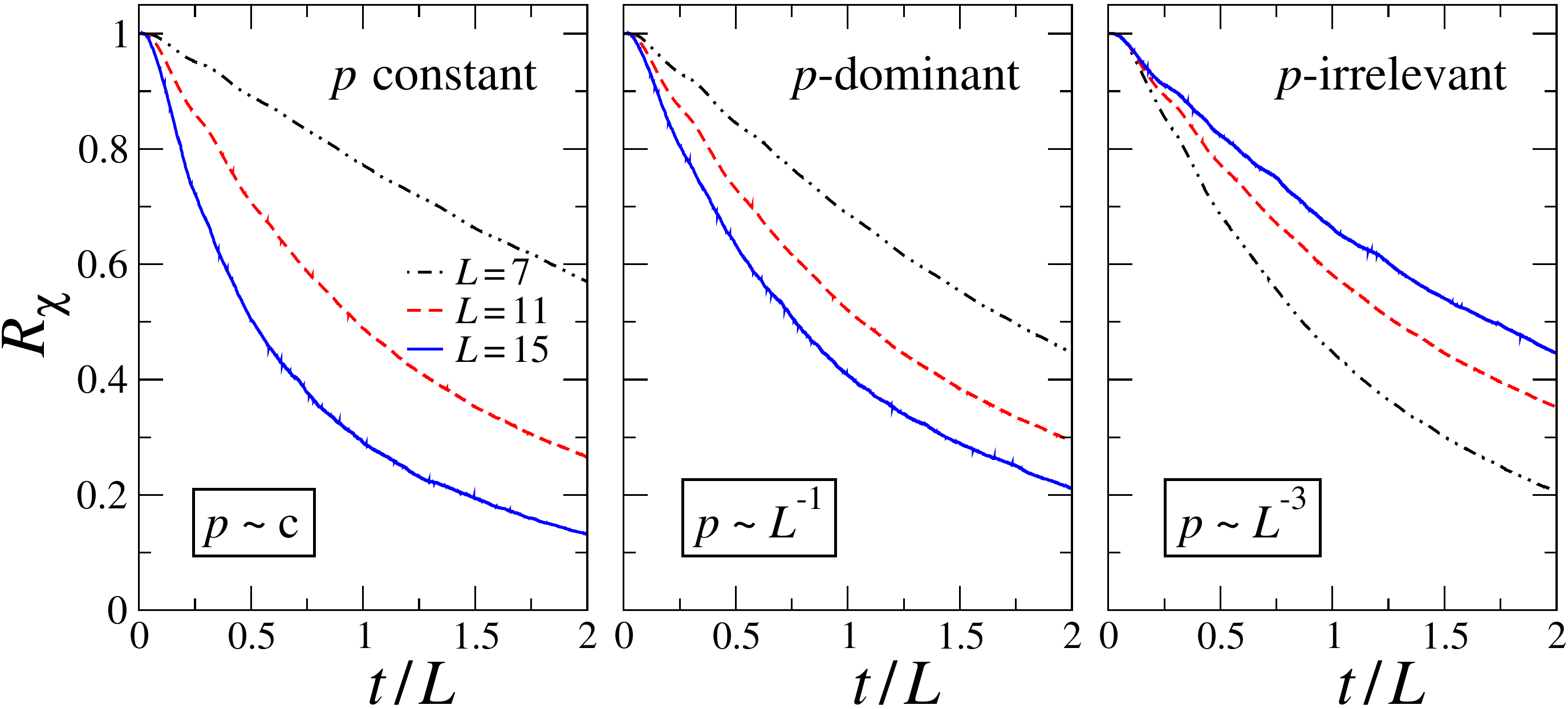}
  \caption{The susceptibility ratio $R_\chi$ vs.~rescaled time $t
    L^{-z}$ for the quantum critical Ising chain ($g=1$, $h=0$) for
    various $L$.  Measurements occur along the longitudinal direction,
    with $t_m = 0.1$ and $p$ being either constant (left: $c=0.005$) or
    equal to $p = c L^{-a}$ (middle: $a=1$, $c=0.05$; right: $a=3$,
    $c=5$).  We note that, in the cases $a=0$ and $a=1$, the time scale $\tau_m$
    of the suppression of the quantum correlations significantly
    decreases with increasing $L$ as a function of the scaling time
    $tL^{-z}$; on the other hand it clearly increases for $a=3$.}
\label{NoScal}
\end{figure}

To achieve a more quantitative understanding of the role of
projective measurements in this context, it is worth focusing on the
quantum critical region, where universality can be helpful to control
the dynamics of the system.  Indeed, the out-of-equilibrium critical
dynamics at continuous quantum transitions develops homogeneous
scaling
laws~\cite{ZDZ-05, Dziarmaga-05, GZHF-10, PSSV-11, CEGS-12, Ulm-etal-13,%%
  Biroli-15, CC-16, PRV-18, PRV-18-lo, NRV-19-wf, RV-19-de}, even in the
presence of dissipation~\cite{YMZ-14, NRV-19-dis}.  One could ask whether
similar scaling arguments hold in the above context, as well.  A naive
application of the dynamic finite-size scaling (FSS)
theory~\cite{Barber-83, Privman-90, CPV-14} at the critical point of
the 1D quantum Ising chain would lead to the results displayed in
Fig.~\ref{NoScal}, where we report the susceptibility ratio $R_\chi$
versus the time scaling variable $t/\tau$ where $\tau=L^{z}\sim \Delta^{-1}$
is the time scale of the critical quantum correlations.
If the probability $p$ to perform measurements is kept constant while
increasing the system size (left panel), the net effect of projections
becomes progressively important, eventually overwhelming the unitary
Hamiltonian dynamics.  Therefore it is clear that a putative scaling
behavior could emerge only after a rescaling of $p$ with $L$.  Guided
by scaling arguments, it is tempting to assume that $p \sim L^{-a}$.
Fig.~\ref{NoScal} shows that, while with $p \sim L^{-1}$ random
measurements are still dominant (middle panel), with $p \sim L^{-3}$
they become irrelevant for the asymptotic dynamic scaling (right
panel).  In between these two cases, there could be a suitable power-law
exponent entering the proper scaling theory for the dynamics arising
from the random measurement protocol described above, provided
this is possible.

Taking advantage of the previous insight, we put forward a
phenomenological scaling theory in which, as working hypothesis, we
assume a scaling behavior for the parameters $t_m$ and $p$
characterizing the measurement procedure.  We
conjecture that a generic observable $B$
(averaged over the trajectories) follows the scaling law
\begin{equation}
  B(\mu,t,t_m,p) \approx  
  b^{-y_B} {\cal B}(\mu b^{y_\mu}, t b^{-z}, t_m \,b^{\zeta},
  p b^{\varepsilon}) \,.
  \label{dynscab}
\end{equation}
Here $b$ denotes an arbitrary positive parameter, $y_{B}$ is the
critical RG dimension of the operator $B$, while $\zeta$ and
$\varepsilon$ are appropriate exponents associated with the
measurement process, and ${\cal B}$ is a universal scaling function
apart from normalizations (more details are provided in
Appendix~\ref{phedynscathe}).  Equation~\eqref{dynscab} is expected
to provide the power-law asymptotic behavior in the large-$b$ limit,
neglecting further dependences on other parameters, which are supposed
to be suppressed (and thus irrelevant) in such limit.

The arbitrariness of the scale parameter $b$ in Eq.~\eqref{dynscab}
can be fixed by setting $b = \lambda \equiv |\mu|^{-1/y_\mu}$, where
$\lambda \sim \xi$ is the length scale of the critical modes.  The
scaling variable associated with the time interval $t_m$ should be
given by the ratio $t_m/\tau$, where $\tau \sim \Delta^{-1}\sim \lambda^z$
is the time scale of the critical models (this implies
$\zeta=-z$).  Keeping $t_m$ fixed in the large-$\lambda$ limit, the
dependence on $t_m$ disappears asymptotically, giving only rise to
$O(\lambda^{-z})$ scaling corrections.  Moreover, noticing that the
parameter $p$ is effectively a probability per unit of time and space,
a reasonable guess would be that its correct scaling to compete with
the critical modes is that $p\sim \lambda^{-z-d}$, thus
\begin{equation}
  \varepsilon = z+d\,.
  \label{kappaguess}
\end{equation}
This leads to the dynamic scaling equation
$B(\mu,t,t_m,p)\approx \lambda^{-y_B} {\cal B}(t \lambda^{-z}, p \lambda^\varepsilon)$~\cite{footnotesiB}.
We stress that the value of
the exponent $\varepsilon$ in Eq.~\eqref{kappaguess} is crucial,
because it allows to separate the measurement-irrelevant regime
$p = o(\lambda^{-\varepsilon})$ (right panel of Fig.~\ref{NoScal}) from the
measurement-dominant regime $p \lambda^{\varepsilon}\to\infty$ (left
and middle panels of Fig.~\ref{NoScal}).  Note that, since $p \sim
\lambda^{-\varepsilon}$ and $t \sim \lambda^z$, the dynamic scaling
ansatz predicts that the time scale $\tau_m$ associated with the
suppression of the quantum correlations behaves as
$\tau_m\sim p^{-\kappa}$ with $\kappa = z/\varepsilon<1$.

\begin{figure}[!t]
  \includegraphics[width=0.98\columnwidth]{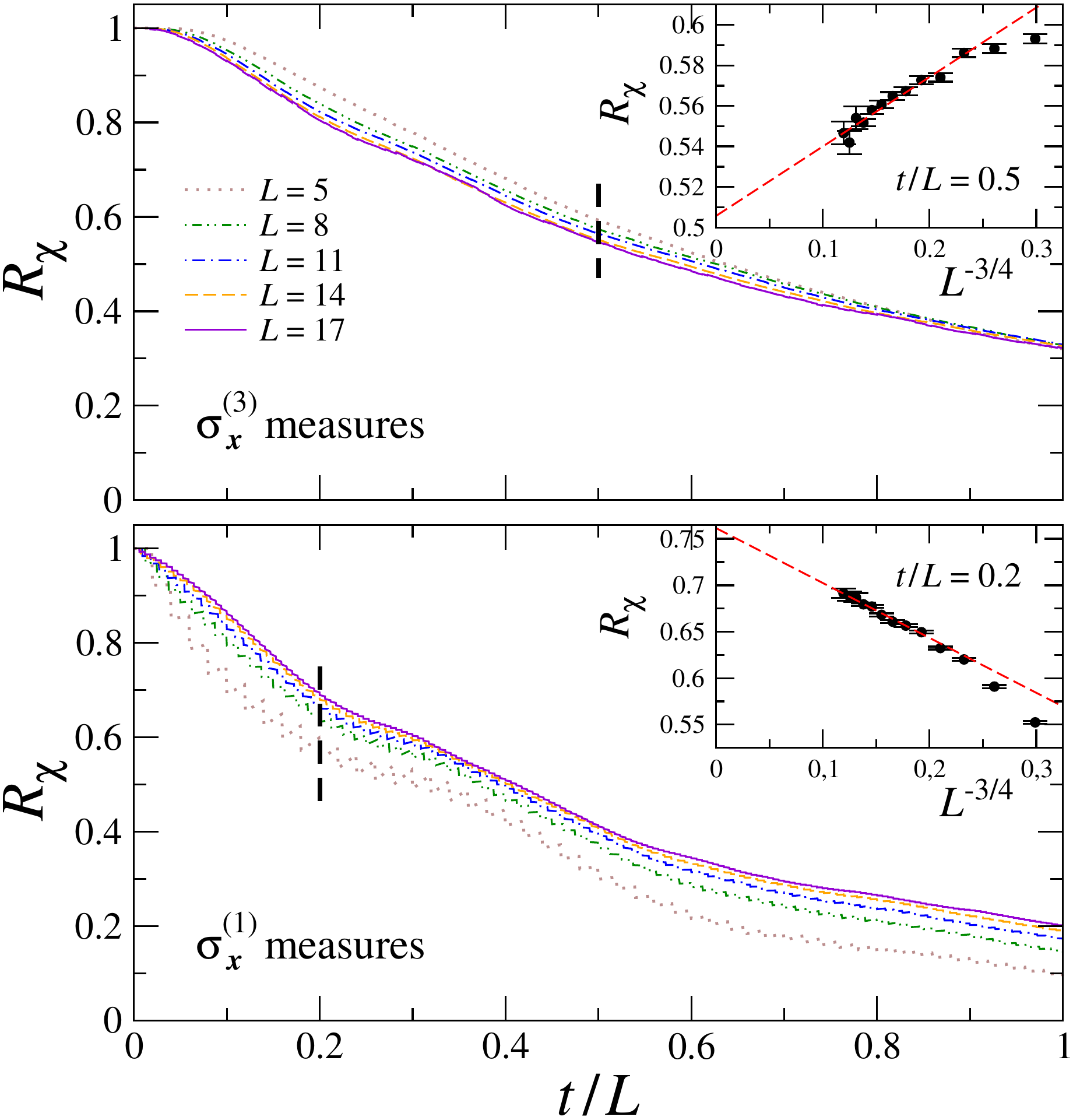}
  \caption{Time behavior of $R_\chi$ at criticality, for
    various sizes.  Random measurements are either along the
    longitudinal (upper plots) or the transverse direction (lower
    plots), for $t_m = 0.1$, $p=1/L^2$ according to the scaling~\eqref{kappaguess}.
    Data converge to an asymptotic scaling function,
    as also shown by the two insets, displaying data for a
    specific value of $t/L$ (dashed line in the main frames).
    They are compatible with a $O(L^{-3/4})$ approach
    to the asymptotic behavior (dashed red lines).
    Analogous scaling results are obtained for $t_m \to 0$,
    as shown in Appendix~\ref{numcomp}.}
\label{fig:susc}
\end{figure}

The above scaling theory holds in the {\em thermodynamic} limit
$L/\lambda\to\infty$, that is expected to be well defined for any
$\mu\neq 0$, for which $\lambda$ is finite.  Nonetheless, for most
practical purposes, both experimental and numerical, one typically
has to face with systems of finite length.  Such situations can be
framed in the FSS framework, where the scale parameter in
Eq.~\eqref{dynscab} is fixed to $b=L$~\cite{Barber-83, Privman-90,
  CPV-14, SGCS-97, CNPV-14, PRV-18}.  Assuming again that $t_m$
is kept fixed, straightforward manipulations lead to the following
scaling law
\begin{equation}
  B(\mu,t,t_m,p,L) \approx  
  L^{-y_B} {\cal B}(\mu L^{y_\mu}, t L^{-z}, p L^{\varepsilon})\,.
  \label{dynfss}
\end{equation}
The proper dynamic FSS behavior is obtained by taking $L\to\infty$,
while keeping $t_m$ and the arguments of the scaling function
${\cal B}$ fixed.

It is worth mentioning that analogous scaling ansatzes for more
general observables, such as fixed-time correlation functions of two
operators, can be obtained using the same arguments and
assumptions (see Appendix~\ref{phedynscathe}).  They can be extended
to include an initial quench of the Hamiltonian parameter $\mu_0 \to \mu$
[by adding a further dependence on $\mu_0 b^{y_\mu}$ in Eq.~(\ref{dynscab})],
to consider finite-temperature initial Gibbs states (by adding a
dependence on $T b^z$), and allowing for weak dissipation~\cite{NRV-19-dis}. 
We note that the scaling arguments do
not depend on the type of local measurement, therefore they
are expected to be somehow independent of them.  Further
investigations are called for to classify the extension of such
independence.

The above phenomenological scaling theory has been checked on the
quantum Ising chain.  The dynamic FSS laws for the magnetization $m$
and its susceptibility $\chi$ follow Eq.~\eqref{dynfss}, in which the
parameter $\mu$ corresponds to either $\delta = g-g_c$ or $h$ in
Eq.~\eqref{hedef}.  In particular, for $\delta = h = 0$, one obtains
$m(t)=0$ by symmetry, and
\begin{equation}
  R_\chi(t,t_m,p,L)\equiv \frac{\chi(t)-1}{\chi(t=0)-1}
  \approx {\cal R}_\chi(t L^{-z},p L^\varepsilon)\,.
  \label{rchiising}
\end{equation}
Further details are provided in Appendix~\ref{dynIsm}.
Results for a system at the quantum critical point, with random local
longitudinal and transverse spin measurements, are shown in
Fig.~\ref{fig:susc}.  The data of $R_\chi$ versus $t L^{-z}$ nicely
agree with Eq.~\eqref{rchiising}.  Corrections to the scaling
are consistent with a $L^{-3/4}$ approach, as expected (see the insets).
An analogous agreement has been obtained for the magnetization at
$h \neq 0$, keeping $h L^{y_h}$ constant (see Appendix~\ref{numcomp}).

\begin{figure}[!t]
  \includegraphics[width=0.95\columnwidth]{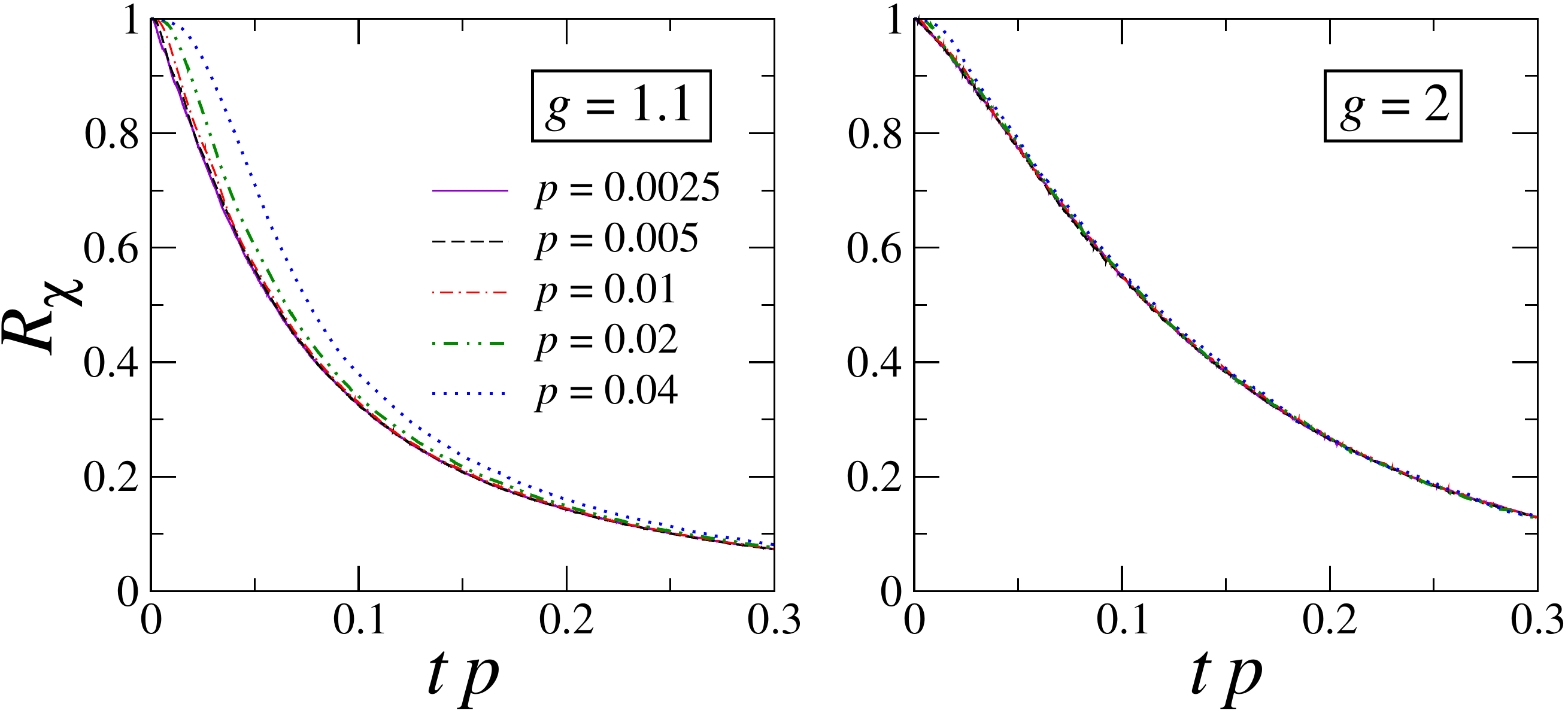}
  \caption{$R_\chi$ for the quantum Ising chain with $L=10$ spins,
    $h=0$, as a function of $t\,p$.  Local measurements are performed
    along $\sigma^{(3)}_x$, with $t_m = 0.1$ and varying $p$ (see
    legend). We fix $g=1.1$ (left) or $g=2$ (right). The approximate
    collapse of the data versus $t\,p$ indicates that, in these
    two cases with $g > g_c$, the characteristic time $\tau_m$
    scales as $\tau_m\sim p^{-1}$.}
\label{fig:NoCrit}
\end{figure}

We finally focus on systems which are not close to a phase transition
(for example, $g > g_c$ in the case of quantum Ising models).
In this case, the system lies
in the disordered phase, where the length scale $\xi$ of the quantum
correlations and the gap $\Delta$ remain finite with increasing $L$.
The data reported in Fig.~\ref{fig:NoCrit} for $R_\chi$ at fixed size
$L$ suggest that, away from criticality, the characteristic time
$\tau_m$ of the measurement process scales as $\tau_m\sim p^{-1}$,
unlike the critical behavior, where 
$\tau_m \sim p^{-\kappa}$ with $\kappa = z/\varepsilon<1$. 

Summarizing, we showed the emergence of different dynamic regimes
arising from the interplay between unitary and projective dynamics in
many-body systems at quantum transitions, where quantum correlations
develop a diverging length scale.  One of them is characterized by the
dominance of the local random measurements, for example for any finite
probability $p$ of making the local measurement.  In contrast, for
sufficiently small values of $p$ (i.e.~decreasing as a sufficiently
large power of the inverse diverging length scale $\xi$), the 
measurements turn out to be irrelevant.  We conjectured  these
two regimes to be separated by dynamic conditions imposing suitable
scaling behaviors for the characterizing parameters of the protocol,
such as the local measurement probability $p$,
which is controlled by the universality class of the quantum
transition.  For a $d$-dimensional critical system, this occurs when
$p\sim \xi^{-\varepsilon}$, for which we argue $\varepsilon=z+d$, where
$z$ is the dynamic exponent of the transition.

This general scenario is supported by numerical results for the quantum Ising chain,
for protocols involving the measurements of either transverse or
longitudinal component of the local spin operator.  Local
measurements generally tend to suppress quantum correlations,
even in the dynamic scaling limit.
The corresponding time scale is expected to behave as
$\tau_m\sim p^{-\kappa}$ with $\kappa=z/\varepsilon<1$, 
to be compared with the noncritical case $\tau_m \sim p^{-1}$.
The smaller power $\kappa$ at the critical point can be explained by 
the fact that the relevant probability ($p_r$) driving the measurement
process is the probability to perform a local measurement within
the critical volume $\xi^d$, therefore $p_r=p\,\xi^d$. The time rate thus
behaves as $\tau_m \sim p_r^{-1}$,
similarly to the noncritical case, where $\xi=O(1)$.

Additional checks are called for, in order to achieve a definite validation of our scaling
conjectures, such as the study of protocols with quenches of the Hamiltonian
parameters, higher dimensions, other quantum transitions
(in particular characterized by different values of the dynamic exponent $z$)
and measurement schemes (non necessarily strictly onsite, but still sufficiently local).
Furthermore we expect that arguments similar to those employed here could be used
for measurements that are localized in restricted regions of space, and also to
derive possible peculiar scaling phenomena in proximity of
first-order quantum transitions, where the boundary conditions could
play a more relevant role~\cite{PRV-18,PRV-18-bc}.

Given the relatively small sizes required to reach
the scaling limit, a direct experimental realization of our protocol can be
reasonably considered as a near-future target for quantum simulations.
Promising platforms are superconducting quantum circuits~\cite{Barends-15, Harris-18, Minev-etal-19},
nuclear spins~\cite{Jiang-etal-09, KCTMHT-16}, trapped ions~\cite{Zhang-17, NMMFH-18, Brydges-19}
and ultracold atomic systems~\cite{BGPFG-09, SWECBK-10, PCV-15, Bernien-17}.

\acknowledgments

We thank M. Collura, A. De Luca, and A. De Pasquale for fruitful discussions.
DR acknowledges the Italian MIUR through PRIN Project No.~2017E44HRF. 

\appendix

\section{Phenomenological scaling theory of the 
out-of-equilibrium dynamics induced by local measurements}
\label{phedynscathe}

We work out a phenomenological scaling theory for the
out-of-equilibrium dynamics arising from random local projective
measurements during the evolution of a many-body system at a quantum
transition~\cite{Sachdev-book}.  For simplicity, we assume that the
quantum transition is driven by a relevant parameter $\mu$ of the
Hamiltonian $H(\mu)$, whose critical value is $\mu_c=0$. At the
critical point, the low-energy unitary Hamiltonian dynamics develops
long-distance correlations, characterized by a diverging length scale
$\xi\sim |\mu|^{-\nu}$, where $\nu=1/y_\mu$ and $y_\mu$ is the
renormalization-group (RG) dimension of the relevant parameter.

More specifically, we consider the dynamic problem associated with
the following protocol: (a) The system starts at $t=0$ from the ground
state close to the critical point, thus $|\mu|\ll 1$; (b) Random local
measurements are performed every time interval $t_m$, with a
homogeneous probability $p$ per site. Between two measurement steps,
the system evolution is driven by the unitary operator $e^{-iH(\mu) t}$.
Hereafter we adopt units of $\hslash = k_B = 1$.

The out-of-equilibrium critical dynamics at continuous
quantum transitions has been shown to obey homogeneous scaling
laws~\cite{ZDZ-05, Dziarmaga-05, GZHF-10, PSSV-11, CEGS-12, Ulm-etal-13,%%
  Biroli-15, CC-16, PRV-18, PRV-18-lo, NRV-19-wf, RV-19-de}, even in the
presence of dissipation~\cite{YMZ-14, NRV-19-dis}.  For example, after an
instantaneous quench from $\mu_0=0$ to $\mu$, a generic observable
$B$ at fixed time $t$ after the quench, is generally expected to behave
as~\cite{PRV-18}
\begin{equation}
  B(\mu,t,L) \approx b^{-y_B} {\cal B}(\mu b^{y_\mu}, t b^{-z}, L/b)\,,
  \label{squsca}
\end{equation}
where $b$ is an arbitrary positive parameter, $L$ is the linear
size of the $d$-dimensional system under investigation,
and ${\cal B}$ is a universal scaling function apart from
normalizations.
The exponent $y_{B}$ denotes the RG dimension of the operator
associated to $B$, while the dynamic exponent $z$ characterizes
the behavior of the energy differences of the lowest-energy states
and, in particular, the ground-state gap $\Delta \sim L^{-z}$.
Equation~\eqref{squsca} is expected to provide the
asymptotic power-law behavior in the large-$b$ limit.

We now extend the dynamic scaling arguments leading to
Eq.~(\ref{squsca}), by allowing for the dependence on the parameters
$t_m$ and $p$ which characterize the measurement procedure of the protocol.
As a working hypothesis, we assume that an asymptotic scaling behavior
is achieved by appropriately rescaling $t_m$ and $p$, such as
\begin{equation}
  B(\mu,t,t_m,p,L) \approx 
  b^{-y_B} {\cal B}(\mu b^{y_\mu}, t b^{-z}, 
  t_m \,b^{\zeta},p b^{\varepsilon}, L/b),\quad
  \label{sdynscab}
\end{equation}
where $\zeta$ and $\varepsilon$ are appropriate exponents
whose relevance is discussed below.

\subsection{Dynamic finite-size scaling}
\label{dynfssca}

It is possible to exploit the arbitrariness of the scale parameter $b$.
For example, by setting $b=L$, we obtain the dynamic
finite-size scaling (FSS) equations, extending those holding for closed
systems~\cite{SGCS-97, CPV-14, CNPV-14, PRV-18}.
Analogously as for the time $t$ that has passed after the quench,
the scaling variable associated with the time interval $t_m$
should be given by the ratio $t_m/\tau$ where $\tau\sim \Delta^{-1}\sim L^z$
is the time scale of the critical models, thus
\begin{equation}
  \zeta=-z\,.  
  \label{szetaval}
\end{equation}
If one keeps $t_m$ fixed in the large-$L$ dynamic FSS limit, the
dependence on $t_m$ disappears asymptotically, giving only rise to
$O(L^{-z})$ scaling corrections.  Moreover, noting that the parameter
$p$ is effectively a probability per unit of time and space, a
reasonable guess is that we must have the power-law scaling behavior
$p\sim L^{-z-d}$ to achieve a nontrivial competition with the critical
modes. Therefore,
\begin{equation}
  \varepsilon = z+d\,.
  \label{skappaguess}
\end{equation}
We stress that the value of the exponent $\varepsilon$ is crucial,
because it allows us to separate the regime $p=o(L^{-\varepsilon})$ in
which the random measurements are irrelevant for the asymptotic
dynamic scaling, from that where they drive the evolution overwhelming
the unitary Hamiltonian dynamics, when $p L^{\varepsilon}\to\infty$.

On the basis of these scaling arguments, from Eq.~\eqref{sdynscab}
we conjecture that, keeping $t_m$ and the arguments of the scaling
function ${\cal B}$ fixed, the dynamic FSS law associated
with the random-measurement protocol reads
\begin{equation}
  B(\mu,t,t_m,p,L) \approx  
  L^{-y_B} {\cal B}(\mu L^{y_\mu}, t L^{-z}, p L^{\varepsilon})\,.
  \label{sdynfss}
\end{equation}
The scaling function ${\cal B}$ is expected to be
largely universal with respect to the Hamiltonian of the system, within
a given universality class, and also with respect to the details
of the protocol. Of course, like any scaling function a quantum
transition, such universality is expected modulo a multiplicative
overall constant and normalizations of the scaling variables.
Note that, in this case, the asymptotic scaling behavior does not
depend on $t_m$ and therefore it is expected to hold also in the limit $t_m\to 0$.

Alternatively, one may rescale the time interval $t_m$ as $\tau\sim L^z$,
thus keeping the ratio $t_m/L^z$ fixed.  In this case we expect the
probability $p$ to scale as the inverse volume only, i.e.
\begin{equation}
  B(\mu,t,t_m,p,L) \approx 
  L^{-y_B} {\cal B}( \mu L^{y_\mu}, t L^{-z},  t_m L^{-z},p L^{d})\,.
  \label{sdFSS2}
\end{equation}
Note that, analogously to Eq.~(\ref{sdynfss}), the FSS limit requires
that $p/t_m \sim L^{-\varepsilon}$.  Similar scaling ansatzes for more
general observables, such as fixed-time correlation functions of two
operators, can be straightforwardly obtained using the same
assumptions and scaling arguments.

The above predictions can be extended to the more complex protocol
including an initial quench of the Hamiltonian parameter from $\mu_0$ to $\mu$;
this is achieved by adding the further dependence on $\mu_0 L^{y_\mu}$.
Moreover, one may also consider an initial Gibbs state for a small
temperature $T$, and this can be taken into account by adding a further
scaling variables $T L^z$.

We finally note that our scaling arguments do not apparently
depend on the type of local measurement, thus they are expected
to be somehow independent on them.

\subsection{Dynamic scaling in the thermodynamic limit}
\label{dynthsca}

To derive a dynamic scaling theory for infinite-volume systems, we may
restart from the general homogeneous power law in Eq.~\eqref{sdynscab}
and set
\begin{equation}
  b = \lambda \equiv |\mu|^{1/y_\mu}\,,
  \label{sbla}
\end{equation}
where $\lambda$ is the length scale of the critical modes, and
consider the limit $L/\lambda\to\infty$, assuming that it is well
defined (this limit corresponds to the so-called {\em thermodynamic}
limit, which is expected to be well defined for any $\mu\neq 0$, for
which $\lambda$ is finite).  Then, keeping again $t_m$ fixed
such that $t_m/\lambda^z\to 0$ in the $\lambda\to\infty$ limit,
and using the fact that the power law associated with $p$ is
expected to be characterized by the same exponent $\varepsilon$ given
in Eq.~(\ref{skappaguess}), one obtains the dynamic scaling ansatz
\begin{equation}
  B(\mu,t,t_m,p)\approx \lambda^{-y_B} {\cal B}(t \lambda^{-z}, p
  \lambda^\varepsilon)\,,
  \label{sthscaB}
\end{equation}
where $\varepsilon=z+d$ is given as in Eq.~\eqref{skappaguess}.
Note that, strictly speaking, one has two scaling functions ${\cal B}$,
depending on the sign of $\mu$.

\section{Dynamic scaling within the quantum Ising model}
\label{dynIsm}

The one-dimensional (1D) quantum Ising model in a transverse field
is one of the simplest paradigmatic quantum many-body systems
exhibiting a nontrivial zero-temperature phase diagram.
The corresponding Hamiltonian reads
\begin{equation}
  H_{\rm Is} = - J \, \sum_{x} \sigma^{(3)}_x \sigma^{(3)}_{x+1} 
  - g\, \sum_x \sigma^{(1)}_x  
  - h \,\sum_x \sigma^{(3)}_x \, ,
  \label{shedef}
\end{equation}
where ${\bm \sigma}\equiv (\sigma^{(1)},\sigma^{(2)},\sigma^{(3)})$
are the spin-1/2 Pauli matrices, the first sum is over all bonds of
the chain connecting nearest-neighbor sites,
while the other sums are over the sites of the chain. 
In our numerical studies, we set $J=1$ as the energy scale
and consider chains of size $L$ with periodic boundary conditions
(${\bm \sigma}_{L+1} \equiv {\bm \sigma}_{1}$).

At $g=g_c$ and $h=0$ (in 1D, $g_c=1$), the model undergoes a continuous quantum
transition belonging to the two-dimensional Ising universality class,
separating a disordered phase ($g>1$) from an ordered ($g<1$) one (see
e.g. Refs.~\cite{SGCS-97, Sachdev-book}). Such transition is
characterized by a diverging length scale $\xi$ of the critical
correlations and the suppression of the ground-state energy gap
$\Delta$ as $\Delta \approx \xi^{-z}$ with $z=1$.
The power-law divergence of $\xi$ is related to the
RG dimensions, $y_\delta$ and $y_h$, of the relevant parameters
$\delta \equiv g-g_c$ and $h$, respectively.
For the transverse field it is given by
\begin{equation}
  y_\delta = 1/\nu \,,
\end{equation}
while for the longitudinal field
\begin{equation}
  y_h = \tfrac12 \big( d+z+2-\eta \big) \,,
\end{equation}
$\eta$ being the exponent which describes the critical behavior
of the correlation function of the order parameter $\sigma^{(3)}$.
Therefore, the critical length scale diverges as $\xi\sim |\delta|^{-1/y_\delta}$
for $h=0$, and $\xi\sim |h|^{-1/y_h}$ for $\delta=0$.
Specializing to the 1D case, one has $d=1$, $\nu=1$, and $\eta = 1/4$,
therefore $y_\delta = 1$ and $y_h=15/8$~\cite{CPV-14}.

For dynamic protocols using the quantum Ising Hamiltonian
in Eq.~\eqref{shedef}, the evolution of the system can be effectively
characterized by the time dependent magnetization along
the coupling direction
\begin{subequations}
\begin{equation}
  M(t) = \frac{1}{L} \sum_{x} \langle \sigma_x^{(3)} \rangle_t\,,
  \label{slongmag}
\end{equation}
the fixed-time longitudinal correlation function
\begin{equation}
  G(x,y,t) = \langle \sigma_x^{(3)} \sigma_y^{(3)}   \rangle_t\,,
  \label{sgxyt}
\end{equation}
and the corresponding susceptibility 
\begin{equation}
  \chi(t) = \frac{1}{L} \sum_{x,y} G(x,y,t)\,.
  \label{schidef}
\end{equation}
\end{subequations}
Here $\langle \cdot \rangle_t$ indicates the expectation value
of a given observable at time $t$.
Note that translation invariance, which also applies in finite-size
systems with periodic boundary conditions, implies $G(x,y,t)\equiv G(x-y,t)$.

For the sake of presentation and without loss of generality, we fix $g=g_c$
and only vary $h$, so that $h$ corresponds to the parameter $\mu$ of the
above-reported scaling equations (analogous equations would hold if $g$ were varied,
with the substitution $h \to \delta$ and $y_h \to y_\delta$).
The dynamic FSS laws of the observables~\eqref{slongmag}-\eqref{schidef},
keeping $t_m$ fixed, thus read
\begin{subequations}
  \begin{eqnarray}
    M(h,t,t_m,p,L) \! & \! \approx \!\! & \! L^{-{y_m}}\, {\cal M}( h L^{y_h}, 
    t L^{-z}, p L^\varepsilon), \qquad \;\;
    \label{sdFSSma}\\
    G(x,h,t,t_m,p,L) \! & \! \approx \!\! & \! L^{-{2y_m}}\, {\cal G}(\tfrac{x}{L}, h L^{y_h}, 
    t L^{-z}, p L^\varepsilon), \qquad \;\;
    \label{sdFSSG}\\
    \chi(h,t,t_m,p,L) \! & \! \approx \!\! & \! L^{d-2y_m} \,{\cal C}(h L^{y_h}, 
    t L^{-z}, p L^\varepsilon), \qquad \;\;
    \label{sdFSSchi}
  \end{eqnarray}
\end{subequations}
where $y_m$ is the RG dimension of the longitudinal spin operator $\sigma_x^{(3)}$,
given by 
\begin{equation}
  y_m = \tfrac12 \big( d+z-2+\eta \big) \,,
  \label{yldef}
\end{equation}
and the power of the prefactor associated with the longitudinal spin
correlation~\eqref{sgxyt} is twice $y_m$
($y_m = 1/8$, in 1D). In particular, for $h=0$ one has $M(t)=0$ and
\begin{equation}
  \chi(t,t_m,p,L) \approx
  L^{d-2y_m} \, {\cal  C}_c(t L^{-z}, p L^\varepsilon)\,.
  \label{sdFSSchi0}
\end{equation}

Corrections to scaling are generally expected to be $O(1/L)$, see for
example Refs.~\cite{CPV-14,PRV-18,RV-19-de}.  However we note that,
in the case of the susceptibility $\chi$ defined as in
Eq.~(\ref{schidef}), $O(L^{-d+2 y_m})$ corrections are also present,
already at the level of the equilibrium ground-state values of $\chi$,
due to analytic contributions to the critical behavior, as explained in Ref.~\cite{CPV-14}.
Therefore, in the case of the Ising chain,
we expect that the leading scaling corrections to the asymptotic dynamic
scaling of the evolution of $\chi$ are $O(L^{-3/4})$.

Our numerical results show that the measurement process generally tends to suppress
quantum correlations, therefore $\chi(t)$ is a monotonic decreasing function.
In particular, the numerics provides evidence of the fact that 
\begin{equation}
  \lim_{t\to\infty} \chi(t) = 1\,.
  \label{chilim}
\end{equation}
The asymptotic value corresponds to a fully disordered state
with vanishing correlations, $G(x,\, y,\, t\!\to\!\infty) = 0$ (for $x \neq y$),
and where the only non-zero contributions entering the sum~\eqref{schidef}
are those for $x=y$, which trivially sum up to one.
To monitor the suppression of quantum correlations due to the
measurement process, it is thus convenient to introduce the ratio
\begin{equation}
  R_\chi = \frac{\chi(t)-1}{\chi(t=0)-1}\,,
  \label{rchidef}
\end{equation}
which goes from one (for $t=0$) to zero (for $t\to\infty$).
In the dynamic scaling limit at the critical point, using Eq.~\eqref{sdFSSchi0},
we can immediately derive the asymptotic behavior
\begin{equation}
  R_\chi(t,t_m,p,L) \approx {\cal R}_\chi(t L^{-z}, p L^\varepsilon)\,.
  \label{sdFSSrchi0}
\end{equation}
Note that, in the dynamic scaling limit, $R_\chi \approx \chi(t)/\chi(0)$,
i.e.~the finite subtraction of one in the numerator and denominator of
the definition of $R_\chi$ turns out to be irrelevant.
Therefore, like for the susceptibility, the approach to the asymptotic
dynamic FSS behavior~\eqref{sdFSSrchi0} is expected to be characterized by
$O(L^{-3/4})$ corrections for the quantum Ising chain (see results
reported in Fig.~\ref{fig:susc}).

In the infinite-volume limit, at $g=g_c$, we expect to have
\begin{subequations}
  \begin{eqnarray}
    M(h,t,t_m,p) &\approx& \lambda^{-y_m} \, {\cal M}
    (t \lambda^{-z},  p \lambda^\varepsilon)\,, \quad \\
    G(x,h,t,t_m,p) &\approx& \lambda^{-2y_m} \, {\cal G}
    (x/\lambda, t \lambda^{-z},  p \lambda^\varepsilon)\,, \quad \\
    \chi(h,t,t_m,p) &\approx& \lambda^{d-2y_m} \, {\cal C}(t \lambda^{-z}, 
    p \lambda^\varepsilon)\,, \quad
  \end{eqnarray}
\end{subequations}
where $\lambda = |h|^{-1/y_h}$.  Such dynamic scaling behaviors are
expected to be approached asymptotically for $L\to\infty$, keeping
fixed the scaling variables of the functions ${\cal M}$ and ${\cal C}$. 

As already noted above, the dynamic scaling arguments that we have
outlined do not apparently depend on the type of local measurement.
In particular, in the case of the quantum Ising model, they should
apply to protocols based on both $\sigma_{\bm x}^{(1)}$ or
$\sigma_{\bm x}^{(3)}$ local measurements.

The time scale $\tau_m$ of the suppression of the quantum correlations
may be estimated from the halving time of $R_\chi(t)$.  
Its power-law scaling behavior  in terms of
the probability $p$ can be easily derived in the dynamic scaling limit,  
by noting that
$p \sim
\xi^{-\varepsilon}$ and $t \sim \xi^z$ where $\xi$ is the length scale
of the critical modes (that is $\xi \sim L$ at the critical point and
$\xi\sim \lambda$ around it).
Therefore, the dynamic scaling  predicts that
the time scale $\tau_m$ associated with the suppression of the quantum
correlations behaves as 
\begin{equation}
  \tau_m\sim p^{-\kappa}\,,\qquad  \kappa = \frac{z}{\varepsilon} = \frac{z}{z+d} \,.
  \label{taump}
\end{equation}
Note that $\kappa<1$, thus the time rate in terms of $p$ turns out to
be accelerated with respect the noncritical behavior $\tau_m\sim p^{-1}$
which has been obtained numerically, see Fig.~\ref{fig:NoCrit}.
This apparently counterintuitive behavior can be explained by 
the nontrivial fact that the relevant probability $p_r$, which drives
the measurement process, is the probability to
perform a local measurement within the critical volume $\xi^d$,
therefore $p_r=p\,\xi^d$. In terms of $p_r$, the time rate thus behaves as
\begin{equation}
  \tau_m \sim p_r^{-1}\,,\qquad  p_r = p \,\xi^d\,,
  \label{taumphat}
\end{equation}
similarly to the noncritical case, where $\xi=O(1)$.

\section{Some details on the numerical computations}
\label{numcomp}

To check our phenomenological dynamic scaling theory discussed before,
we have performed some numerical simulations on the 1D quantum
Ising chain~\eqref{shedef}, based on exact diagonalization (ED).
We are interested in the random-measurement protocol starting from
the ground state of a system of size $L$ (with periodic boundary
conditions) for the Hamiltonian parameter $h$ and with $g = g_c = 1$,
which has been obtained by means of a Lanczos technique.
The evolution, monitored through a fourth-order Suzuki-Trotter
decomposition of the unitary-evolution operator with time step $dt = 0.005$,
is essentially driven by the random measurements, which are performed
at every time interval $t_m$. We have considered either local longitudinal
[$\sigma_x^{(3)}$] or transverse [$\sigma_x^{(1)}$] measurements,
occurring with a probability $p$ per site. 

\begin{figure}[!t]
  \includegraphics[width=0.98\columnwidth]{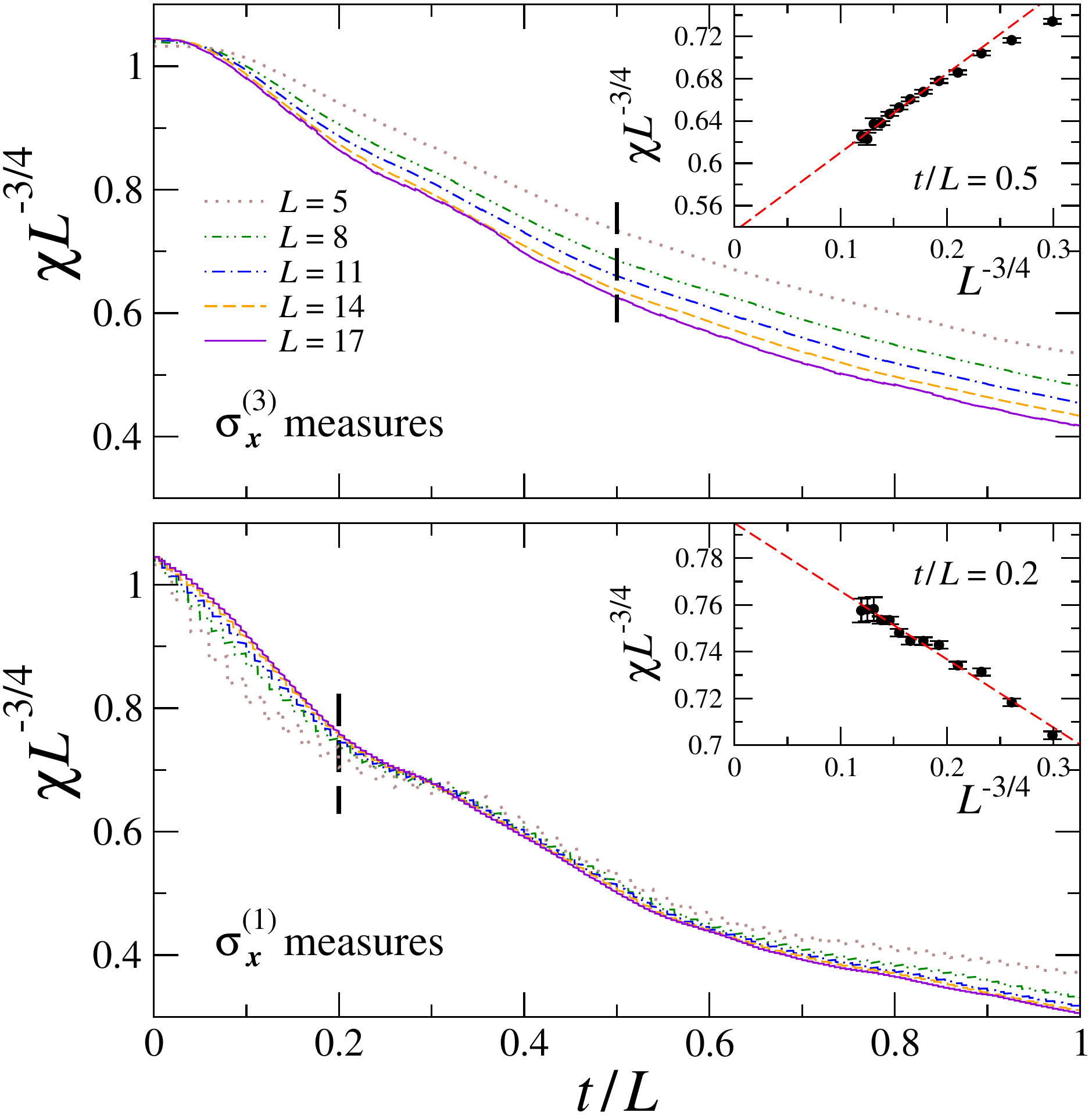}
  \caption{Rescaled susceptibility $\chi L^{-3/4}$ at criticality
    ($g=1$, $h=0$), as a function of the rescaled time $t/L$,
    for the quantum Ising chain with different system sizes $L$ (see legend).
    The upper frame refers to random measures along the longitudinal
    direction [$\sigma^{(3)}_x$], while the lower frame refers to random
    measures along the transverse direction [$\sigma^{(1)}_x$].
    Measurements are characterized by $t_m = 0.1$, $p=1/L^2$ as
    predicted by Eq.~\eqref{skappaguess}.  The two insets display the
    same data of the main panels for a specific cut in the rescaled
    time ($t/L = 0.5$ for the upper frame, $t/L = 0.2$ for the lower
    frame) versus $L^{-3/4}$.  The approach to the asymptotic value
    is consistent with the expected $O(L^{-3/4})$ corrections.
    Dashed red lines denote $1/L^{3/4}$ fits to numerical data (black circles)
    and have been obtained by discarding points for the smaller available sizes.
    Here and in the next figures, data have been averaged over a number of
    $N_{\rm avg} = 2 \times 10^3$ (for $L \leq 14$) and 
    $N_{\rm avg} = 2 \times 10^3$ (for $L \geq 15$) trajectories.}
  \label{fig:chi}
\end{figure}

In Fig.~\ref{fig:susc} we showed results only for the susceptibility ratio $R_\chi$.
Here we provide some additional data, both for
the magnetization~\eqref{slongmag} and for the susceptibility~\eqref{schidef},
in which we kept $g=1$ and varied the longitudinal field $h$
(note that the quantum Ising chain with $h\neq 0$ is not integrable).

\begin{figure}[!t]
  \includegraphics[width=0.98\columnwidth]{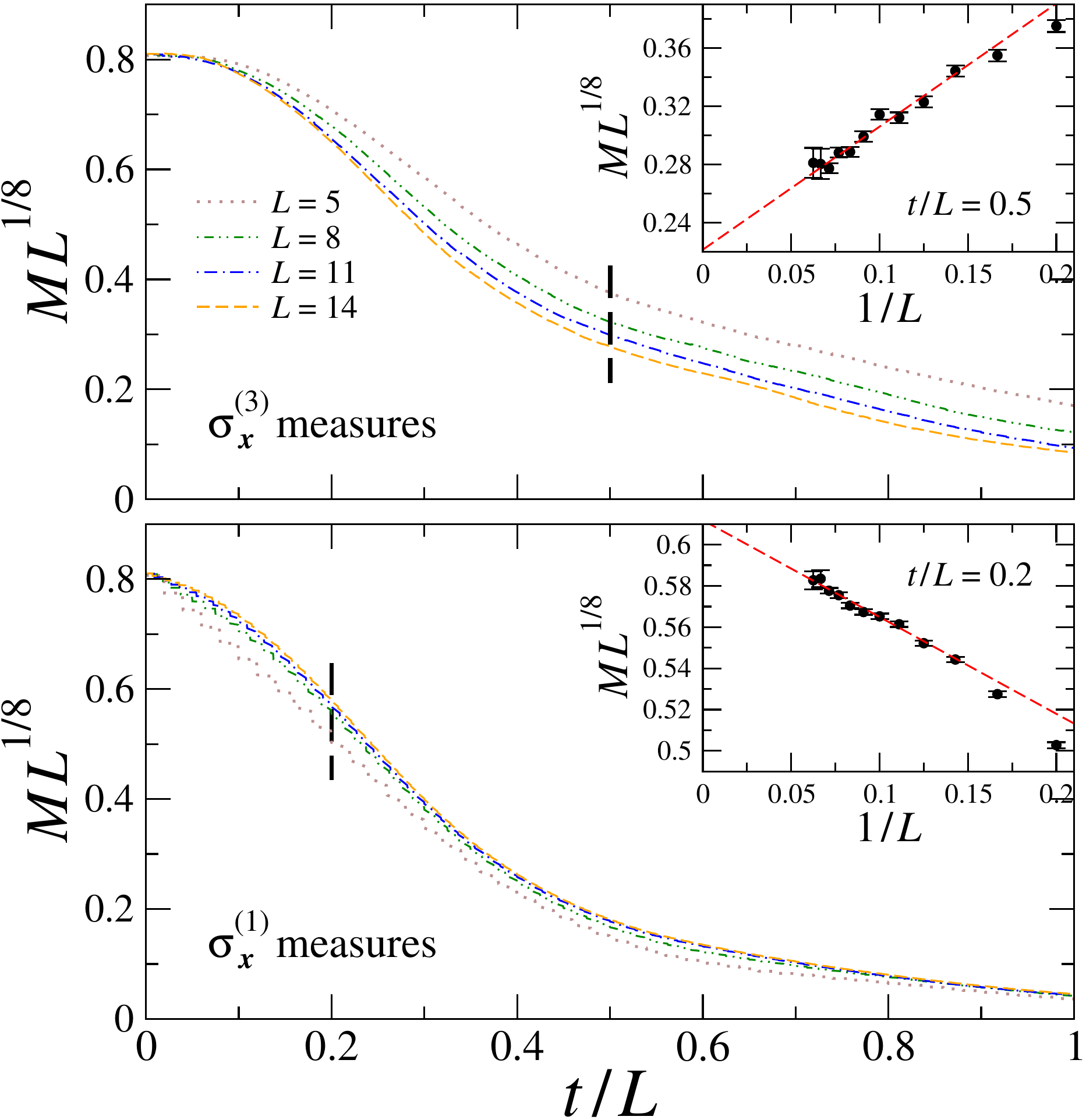}
  \caption{Same as in Fig.~\ref{fig:chi}, but for the rescaled
    magnetization $M L^{1/8}$ vs.~the rescaled time $t/L$.
    The initial state here is the ground state of the quantum Ising ring with
    $g=1$ and rescaled longitudinal field $h L^{15/8} = 1$, which has a
    finite magnetization.  The two insets display the same data of the
    main panels for a specific cut in the rescaled time, versus $1/L$.
    The approach to the asymptotic FSS behavior, for finite $L$, appears
    to be consistent with the expected $O(1/L)$ trend (dashed red lines).}
\label{fig:magn}
\end{figure}

The need of averaging over many different trajectories, typically $O(10^4)$,
together with the fact that the numerical results shown in this paper
are nicely consistent with the dynamic FSS theory, prevented us from
studying systems with more than $L=18$ sites, although larger sizes
would be easily addressable for a single trajectory or a few ones.
Also note that we preferred to use conventional (and fully controllable)
ED techniques over DMRG-based algorithms~\cite{Schollwock-11, Daley-14},
since with those latter methods it is more complicated to guarantee the required
accuracy in order to carefully test our phenomenological scaling theory.
Nonetheless, there are no conceptual limitations in using DMRG 
for analyzing the measurement-induced dynamics of quantum lattice
models with finite degrees of freedom~\cite{TZ-19}. 
In summary, ED techniques are more controllable, but suffer from severe
limitations in the reachable system sizes; DMRG allows to study 
larger systems, although it requires more care in the choice of the bond-link
dimension for the study of dynamical problems.

Figure~\ref{fig:chi} displays the numerical outcomes for the susceptibility
at the Ising critical point ($g=1$, $h=0$), for random local measurements
taken along the longitudinal or the transverse direction.
Note that the results presented in this figure are the same as those reported
in Fig.~\ref{fig:susc}, but for the rescaled susceptibility
$\chi \, L^{-3/4}$ [instead of the ratio $R_\chi$ in Eq.~\eqref{rchidef}].
Similarly as for the susceptibility ratio, we observe a nice agreement with the predicted
scaling behavior in Eq.~\eqref{sdFSSchi}. Moreover, corrections to the scaling
are consistent with a $L^{-3/4}$ behavior, as expected (see the two insets).

Results for the magnetization $m(t)$ are reported in Fig.~\ref{fig:magn}.
In that case, we considered $g=1$ and a nonzero longitudinal field $h$, since
the latter is essential in order to start from an initially magnetized state [$m(0) \neq 0$].
After a suitable rescaling of all the relevant parameters, the various curves
approach an asymptotic scaling behavior, as indicated in Eq.~\eqref{sdFSSma}.
Notice that we also rescaled the field $h$ so to keep the scaling variable
$h L^{15/8}$ constant.
The approach to the scaling is governed by corrections whose leading order
appear to be consistent with a $L^{-1}$ behavior, as witnessed by the two insets.

\begin{figure}[!t]
  \includegraphics[width=0.98\columnwidth]{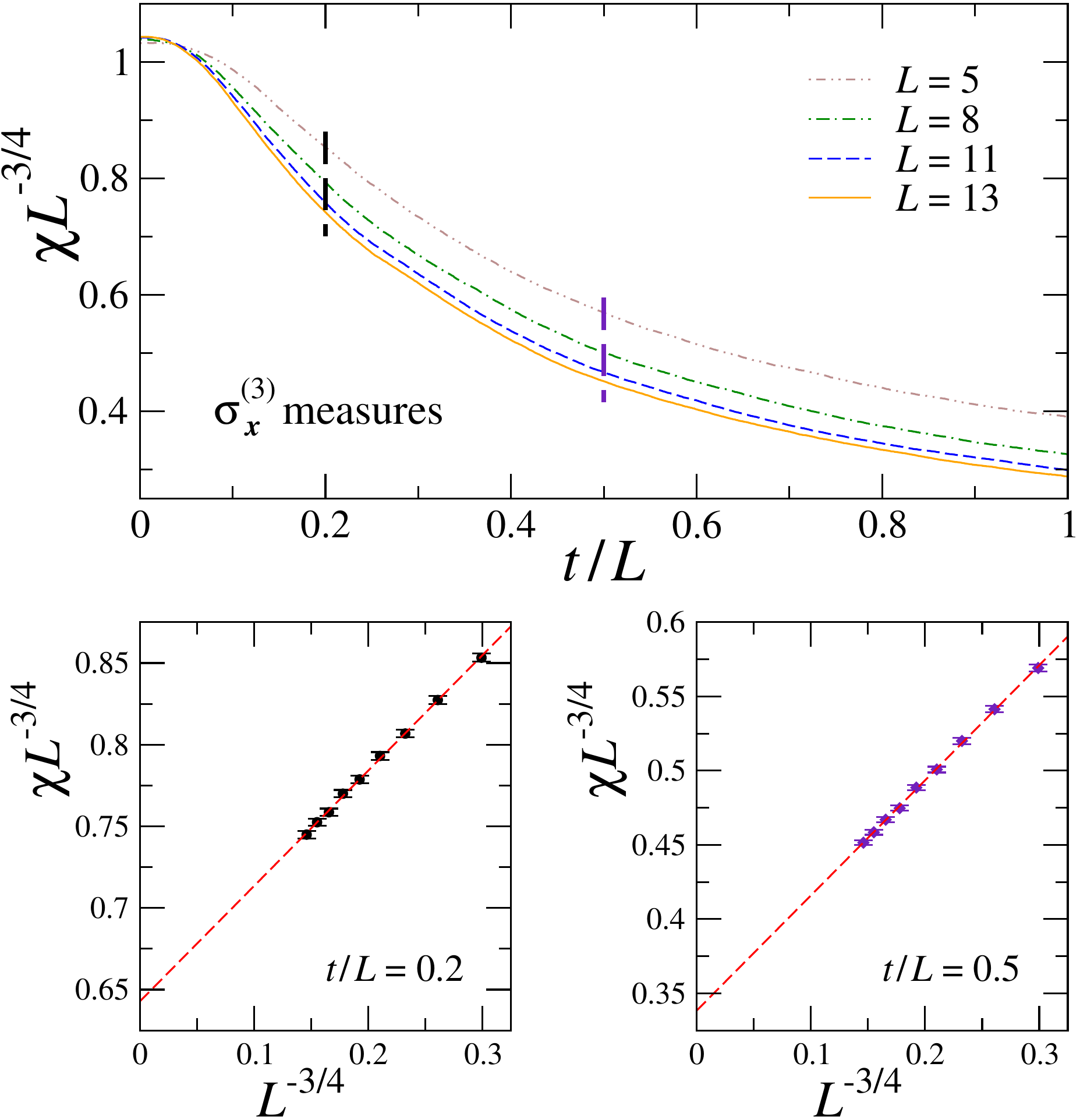}
  \caption{Same as in Fig.~\ref{fig:chi}, but for random local measurements
    along the longitudinal direction in the continuous time limit $t_m \to 0$
    and with $p = 0.1/L^2$. The lower panels display the same data of the main panel,
    for two specific values of rescaled time (left: $t/L = 0.2$; right: $t/L = 0.5$)
    versus $L^{-3/4}$. The approach to the asymptotic value is consistent with the
    expected $L^{-3/4}$ corrections (dashed red lines).}
\label{fig:chi_cont}
\end{figure}

All the numerical data presented above correspond
to fixing the time interval between two consecutive measurements equal to $t_m = 0.1$.
We have checked that analogous scaling results can be obtained for arbitrary
values of $t_m$. In particular, in Fig.~\ref{fig:chi_cont} we have considered
the limit $t_m \to 0$. More precisely, random local measurements have been performed
at every Trotter time step, so that $t_m = 0.005$.
The upper panel displays results for the rescaled susceptibility $\chi L^{-3/4}$
as a function of the rescaled time $t/L$, keeping $g=1$ and $h=0$,
for measurements performed along the longitudinal direction [$\sigma_x^{(3)}$].
The lower panels highlight that corrections to the scaling are $O(L^{-3/4})$, as expected.
Notice that here the compatibility with a $L^{-3/4}$ behavior (dashed red lines) is excellent
already at very small system sizes ($L=5$), contrary to the case of larger $t_m$ values:
compare with the insets of Fig.~\ref{fig:chi}, where deviations from the expected
trend emerge at smaller $L$.
This hints at the fact that other subleading terms, that may enter the scaling corrections
at finite $L$, may get suppressed in the limit $t_m \to 0$, such as those which are $O(t_m / L^z)$.

Finally we observe that the same scaling functions (${\cal M}$, ${\cal G}$, ${\cal C}$, \ldots)
are expected to hold for vanishing $t_m$,
so that the asymptotic curves for the magnetization and the susceptibility at $L \to \infty$
should coincide with those at finite $t_m$, after a proper rescaling of all the relevant
parameters in the dynamic protocol.

\section{One-spin model subject to periodic measurements}
\label{onespin}

Here we discuss the dynamics of a single spin-$1/2$ system
in a magnetic field, subject to periodic measurements along
a given axis, for which it is possible to derive an analytic solution.
We consider the following Hamiltonian model:
\begin{equation}
  H_1 = - g\, \sigma^{(1)} - h \,\sigma^{(3)} \,,
  \label{onespinh}
\end{equation}
where $g$ and $h$ denote the intensity of an applied
external magnetic field along two orthogonal directions
[${\bm \sigma}\equiv (\sigma^{(1)},\sigma^{(2)},\sigma^{(3)})$
are the usual spin-1/2 Pauli matrices].
Without loss of generality, we fix $g=1$.  We now suppose to initialize
the system in its ground state, associated with the
parameter $h$ at $t=0$. Then we perform a sequence of repeated measurements
of the operator $\sigma^{(3)}$ at every time interval $t_m$
(the choice of the measurement axis is arbitrary).

The dynamics arising from this protocol can be described in terms of
the system's density matrix $\rho$. The starting ($t=0$) state is a pure state,
given by
\begin{equation}
  \rho_0\equiv \rho(t=0) = |0_h\rangle \langle 0_h |\,,
  \label{startingpoint}
\end{equation}
where 
\begin{eqnarray}
  E_0 & = & - \sqrt{1 + h^2} \,,\label{loweig}\\
  |0_h \rangle & = & {\cal N} \left[
    (-h - \sqrt{1 + h^2})|+\rangle + |-\rangle \right]\,,
  \label{loweigs}
\end{eqnarray}
and $|\pm\rangle$ are the eigenstates of $\sigma^{(3)}$,
while ${\cal N}$ is the normalization to obtain $\langle 0_h|0_h\rangle=1$.
Then, defining the density matrix after $n$ measurements, at time $t=nt_m$, as
\begin{equation}
  \rho_n \equiv \rho(t=nt_m)\,,
  \label{rhondef}
\end{equation}
the subsequent dynamics can be described as a series of two-step
operations: \\
{\it (i)} A unitary time evolution for a time $t_m$,
\begin{subequations}
\begin{equation}
  \tilde{\rho}_{n+1} = e^{-iH_1t_m} \rho_n e^{iH_1t_m}\,;\label{firststep}
\end{equation}
{\it (ii)} The measurement of $\sigma^{(3)}$, 
\begin{equation}
  \rho_{n+1} = {\rm Tr}(\tilde{\rho}_{n+1} P_+) \,P_+ + 
      {\rm Tr}(\tilde{\rho}_{n+1} P_-) \,P_-\,,\label{measstep}
\end{equation}
where $P_\pm$ are the projectors onto the eigenstates $|\pm\rangle$ of
$\sigma^{(3)}$.
Simple manipulations thus lead to
\begin{equation}
  \rho_{n+1} = \tfrac12 I + W_{n+1} \sigma^{(3)},\qquad 
  W_{n+1} = \tilde{\rho}_{n+1}^{11} - \tfrac12 \,,
  \label{measstepB}
\end{equation}
\end{subequations}
where $W$ quantifies the deviation from the trivial
completely unpolarized density matrix $\rho_u=I/2$.  Given the initial
condition (\ref{startingpoint}), one finds
\begin{equation}
  W_1 = \frac{h (\sqrt{1+h^2} + h)}{2(1 + h^2 + h \sqrt{1+h^2})}\,.
  \label{w1comp}
\end{equation}

Straightforward computations allow to obtain
\begin{equation}
  W_{n+1} = f \, W_n\,,\qquad
  f = 1 - \frac{2 [\sin(t_m \sqrt{1+h^2})]^2}{1 + h^2} \,.
  \label{wnsequ}
\end{equation}
In particular, for $h=1$, one gets
$W_1= (1 + \sqrt{2})/(4 + 2 \sqrt{2})\approx 0.35355$
and $f = 1 - \sin^2(\sqrt{2}t_m)$.
Note that the factor $f$ is bounded, indeed
\begin{equation}
  |f|\le 1\,.
  \label{bf}
\end{equation}
Moreover, for arbitrary values of $h$ and $t_m$, one strictly finds $|f|<1$.
Indeed $f=-1$ for $h=0$ only, while $f=1$ for the specific values
$t_m = m\pi/\sqrt{1+k^2}$, with integer $m$.

Equation~\eqref{wnsequ} implies 
\begin{equation}
  W_n = W_1 f^{n-1} \,.
  \label{wnfn}
\end{equation}
Therefore, by monitoring the expectation value of $\sigma^{(3)}$,
one eventually gets
\begin{equation}
  \langle \sigma^{(3)} \rangle_{t=nt_m}
  = {\rm Tr}[\sigma^{(3)} \rho_n]  = 2 W_n=2 W_1 f^{n-1}\,.
  \label{finresones}
\end{equation}
Since in general $|f|<1$, this shows that for any $h$ the dynamic
protocol tends to produce disorder the spin model, leading to a completely
unpolarized matrix density.

For the specific case of sufficiently small $t_m \ll 1$, one finds
\begin{equation}
  f = 1 - 2 t_m^2 + O(t_m^4)\,,
  \label{fexp}
\end{equation}
and thus
\begin{equation}
  W_n \sim (1 - 2 t_m^2)^n \approx e^{-2 t_m^2 n }\,.
  \label{wnasy}
\end{equation}
Therefore, the dependence on $h$ disappears in the leading $O(t_m^2)$ term.
Finally we note that, in the limit $t_m\to 0$, the quantum Zeno
effect~\cite{MS-77, FP-08} can be recovered,
indeed one simply has $f \to 1$ and $W_n\to 1$.


\begin{thebibliography}{99}
\bibitem{Ising1D} For the 1D quantum Ising model, the critical point
  is located at $g_c=1$, $h=0$. The RG dimension of the two control
  parameters $\delta$ and $h$ are, respectively,
  $y_\delta = 1$ and $y_h=15/8$.  The exponent describing the critical
  two-point function is $\eta=1/4$, which determines the RG
  dimension of the longitudinal spin operator $y_m = (d+z-2+\eta)/2$
  and the power law $\chi \sim \xi^{d-2y_m}$~\cite{CPV-14}.
  For 2D models, the
  critical exponents are not known exactly, but there are very
  accurate estimates~\cite{PV-02}, and in particular~\cite{KPSV-16}:
  $y_\delta=1/\nu$ with $\nu=0.629971(4)$ and $y_h = (5-\eta)/2$ with
  $\eta=0.036298(2)$.  For 3D systems, they assume mean-field values,
  $y_\delta=2$ and $y_h=3$, apart from logarithmic corrections.
\bibitem{Numerics} The Ising ground state has been obtained through
  Lanczos diagonalization, while for the dynamics between two
  consecutive measurements we employed a fourth-order Suzuki-Trotter
  decomposition of the unitary-evolution operator, with time step $dt
  = 5 \times 10^{-3}$ (we checked that this value is sufficient to
  ensure convergence, on the scale of all the figures, over the
  largest reached system size).  Results at any time $t$ have been averaged over
  $N_{\rm avg}$ different trajectories, with $N_{\rm avg}= 2 \times 10^4$
  up to $L=14$ sites, and $N_{\rm avg} = 2 \times 10^3$ for $L > 14$.
\bibitem{footnotesiB}
  Note that, strictly speaking, one has two scaling
  functions ${\cal B}$, depending on the sign of $\mu$.
\end{thebibliography}
\end{document}